\documentclass[12pt,a4paper]{article}

\usepackage[utf8]{inputenc}
\usepackage[T1]{fontenc}
\usepackage[english]{babel}
\usepackage{lmodern}
\usepackage{amsmath,amssymb,amsfonts}
\usepackage{bm}
\usepackage{geometry}
\usepackage{setspace}
\usepackage{cite}
\usepackage{hyperref}

\hypersetup{hidelinks}

\title{Deformation Quantization of Schutz Quantum Cosmology:\\
Relational Time, Constraints, and Physical States}

\author{Raphael Fracalossi\thanks{E-mail:
  \href{mailto:raphael.fracalossi@ufop.edu.br}{raphael.fracalossi@ufop.edu.br}}\\[2mm]
  {\small Departamento de F\'isica, Instituto de Ci\^encias Exatas e Biol\'ogicas,}\\
  {\small Universidade Federal de Ouro Preto, 35402-136 Ouro Preto, MG, Brazil}}

\date{}

\begin{document}

\onehalfspacing
\raggedbottom

\maketitle

\begin{abstract}
A controlled correspondence is established between reduced relational
dynamics and extended constrained quantization in an FLRW cosmology
coupled to a Schutz perfect fluid. A canonical transformation in the
matter sector yields a deparametrized constraint of the form
$\mathcal C=p_T+H$, with $T$ serving as an internal time. In the extended
Weyl--Wigner formulation, the quantum constraint is imposed bilaterally.
Its antisymmetric part gives the exact Moyal evolution generated by the
reduced Hamiltonian, while its symmetric part imposes an independent
quantum shell condition. For a fixed self-adjoint realization of $\hat
H$, every reduced density operator supported on its negative spectral
subspace admits an explicit distributional extension. A spectral dyad
with energies $E$ and $E'$ is mapped to a clock Wigner distribution
supported at $p_T=-(E+E')/2$, with relational phase determined by
$E-E'$. The map preserves off-diagonal coherences, satisfies both star
constraints, and has the reduced Wigner state as its $p_T$ marginal; it
is therefore injective on the admissible positive-density sector. A flat
dust model with $q=1/2$ and Dirichlet data provides an explicit example:
its reduced spectrum is $(-\infty,0]$, so the admissible sector contains
all normalizable reduced states. The correspondence applies to a fixed
constraint representative and quantum realization.
\end{abstract}


\section{Introduction}

The construction of a quantum theory of gravity requires a reformulation
of the dynamical structure of general relativity, in which the
gravitational field is no longer treated as a fixed background but becomes
a dynamical degree of freedom. In the canonical formulation of general
relativity, this feature leads naturally to a constrained Hamiltonian
system, where the dynamics is generated by constraints rather than by an
ordinary Hamiltonian evolution. Following the Dirac approach to
constrained systems, the identification of physical states and
observables requires a careful treatment of these constraints
\cite{Dirac1950,Dirac1964,ArnowittDeserMisner1962,HenneauxTeitelboim1992}.

Quantum cosmology provides a natural framework in which these conceptual
issues can be investigated in a simplified setting. By restricting the
gravitational degrees of freedom to homogeneous and isotropic
configurations, minisuperspace models reduce the infinite-dimensional
phase space of general relativity to a finite-dimensional system while
preserving the essential structure of the Hamiltonian constraint.
Despite their simplicity, such models have played an important role in
the study of quantum gravitational dynamics, singularity avoidance, and
the emergence of semiclassical cosmological behavior
\cite{Halliwell1991,Kiefer2012}.

In canonical quantum cosmology, the Hamiltonian constraint gives rise to
the Wheeler--DeWitt equation,

\[
\hat{\mathcal H}\Psi=0,
\]
as originally formulated in canonical quantum gravity
\cite{DeWitt1967}. It differs fundamentally from the Schrödinger equation because the
external time parameter of ordinary quantum mechanics is absent. This
feature is closely related to the problem of time in quantum gravity: if
time is not an external structure but part of the dynamical geometry,
the meaning of quantum evolution must be reformulated in terms of
correlations among physical degrees of freedom
\cite{Kuchar1992,Isham1993,Kiefer2012}.

One possible strategy to address this issue consists of introducing
internal degrees of freedom that can play the role of relational clocks.
Matter fields and perfect fluids have been extensively investigated in
quantum cosmology as possible sources of such internal temporal
variables. Among the different approaches, the Schutz formalism provides
a particularly useful framework because it allows the canonical
description of perfect fluids and, after suitable transformations,
introduces matter variables that may appear linearly in the Hamiltonian
constraint \cite{Schutz1970,Schutz1971}.

This property has motivated several applications of Schutz variables in
quantum cosmology. When the momentum associated with the fluid variable
enters the constraint linearly, the corresponding variable can be
interpreted as an internal time parameter, allowing the construction of a
reduced relational description. In this approach, the quantum dynamics
resembles a Schrödinger evolution with respect to a degree of freedom of
the system itself rather than with respect to an external temporal
parameter \cite{AlvarengaEtAl2002,MoneratEtAl2020}.

However, the use of relational time also raises a fundamental question
concerning the relation between reduction and quantization. In
constrained systems, one may either solve the classical constraint and
quantize the reduced system or preserve the extended constrained phase
space and impose the quantum constraint after quantization. Although
these procedures may coincide in particular situations, their
compatibility is not guaranteed in general. This issue becomes
especially relevant when the quantum structure of phase space is
formulated through noncommutative products.

Deformation quantization, originally developed through the phase-space
formulations of Weyl, Wigner, Groenewold, and Moyal
\cite{Weyl1931,Wigner1932,Groenewold1946,Moyal1949}, provides an
alternative formulation of quantum mechanics in which quantum states are
represented directly as functions on phase space and operator products
are replaced by the noncommutative star product. The formalism was later
systematized in a broader mathematical framework
\cite{Bayen1978I,Bayen1978II} and has been applied to different classes
of quantum systems \cite{Hillery1984,CurtrightFairlieZachos2014}.

For constrained systems, deformation quantization offers a natural
language in which the Hamiltonian constraint and the quantum state
coexist in the same phase-space formulation. In particular, the Weyl
symbol of the constraint may act from both sides of the phase-space
state, leading to the bilateral conditions
\[
\mathcal C_W\star W=0,
\]
and
\[
W\star\mathcal C_W=0.
\]

These equations provide a deformation-quantized realization of the
quantum constraint and raise the question of how they relate to the
relational dynamics obtained after reduction. Phase-space quantization
has also been applied directly to cosmological models
\cite{CorderoGarciaCompeanTurrubiates2011,RashkiJalalzadeh2017}, and,
for constrained systems, physical Wigner distributions have been
constructed through a star-product version of group averaging
\cite{BerraMontielMolgado2020}.
Recent work has additionally used deformation quantization for a test
particle propagating on a prescribed FLRW background
\cite{BobadillaCembranos2025}. That problem differs from the one studied
here: the scale factor is dynamical in the present minisuperspace model, the
Schutz fluid supplies an internal clock, and the phase-space state is
subject to a Hamiltonian constraint.

This problem is investigated here in the context of
deformation-quantized Schutz quantum cosmology. Starting from the
reduced relational formulation, in which the fluid variable provides an
internal clock, the analysis identifies sufficient conditions under which the reduced
dynamics admits a consistent reconstruction in the extended
phase-space formulation with bilateral quantum constraints.

The reconstruction associates admissible reduced relational states with extended states
satisfying the bilateral constraint while preserving the relational
dynamics generated by the reduced Hamiltonian. Its claim is deliberately
restricted to compatibility within that sector; no universal equivalence
between reduced and extended quantization is assumed.

The paper is organized as follows. Section~\ref{sec:schutz} introduces
the Schutz quantum cosmological framework, the reduced relational
formulation, and an explicit flat dust realization.
Section~\ref{sec:deformation-constraints} presents the
deformation-quantization description of constrained systems, and
Sec.~\ref{sec:reconstruction} analyzes the relation between reduced and
extended formulations. Section~\ref{sec:discussion} discusses the
physical interpretation and scope of the reconstructed sector, and
Sec.~\ref{sec:conclusion} summarizes the results.

\section{Schutz quantum cosmology and relational dynamics}
\label{sec:schutz}

\subsection{Minisuperspace dynamics and the Schutz clock}
\label{subsec:schutz-clock}

Consider a homogeneous and isotropic spacetime described by the FLRW
line element \cite{Halliwell1991,Kiefer2012}

\begin{equation}
 ds^2=-N^2(t)\,dt^2+a^2(t)\,d\Sigma_k^2,
 \label{eq:flrw-metric}
\end{equation}
where $N(t)$ is the lapse function, $a(t)>0$ is the scale factor, and
$k=0,\pm1$ denotes the spatial curvature. After integration over the
homogeneous spatial section and absorption of the corresponding
fiducial volume, the gravitational Lagrangian may be written, in units
$16\pi G=1$, as

\begin{equation}
 L_g=-\frac{6a\dot a^2}{N}+6kNa.
 \label{eq:grav-lagrangian}
\end{equation}

The momentum conjugate to the scale factor is

\begin{equation}
 p_a=-\frac{12a\dot a}{N},
 \label{eq:pa-definition}
\end{equation}
and the gravitational contribution to the super-Hamiltonian is therefore

\begin{equation}
 \mathcal H_g=-\frac{p_a^2}{24a}-6ka.
 \label{eq:grav-super-hamiltonian}
\end{equation}

The lapse has no independent velocity and acts as a Lagrange multiplier.
Thus, despite the minisuperspace truncation, the cosmological dynamics
retains the constrained character inherited from general relativity.

The matter sector is a Schutz perfect fluid with barotropic equation of
state \cite{Schutz1970,Schutz1971}

\begin{equation}
 p=\alpha\rho,
 \qquad
 \gamma=1+\alpha.
 \label{eq:barotropic-eos}
\end{equation}

For homogeneous and irrotational configurations, the relevant fluid
degrees of freedom can be represented by the canonical pairs
$(\epsilon,p_\epsilon)$ and $(s,p_s)$, where $s$ is the specific entropy.
For the reconstruction problem, the relevant feature of this formulation
is that a canonical transformation makes the fluid Hamiltonian linear in
one momentum. On the regular matter chart $p_\epsilon>0$, the clock variables are
defined as
\cite{AlvarengaEtAl2002}

\begin{equation}
 T=-p_s e^{-s}p_\epsilon^{-\gamma},
 \qquad
 p_T=e^s p_\epsilon^\gamma,
 \label{eq:schutz-clock-transformation}
\end{equation}
together with the complementary variables

\begin{equation}
 \bar\epsilon=\epsilon+\gamma\frac{p_s}{p_\epsilon},
 \qquad
 \bar p_\epsilon=p_\epsilon.
 \label{eq:schutz-complementary-pair}
\end{equation}

They satisfy

\begin{equation}
 \{T,p_T\}=1,
 \qquad
 \{\bar\epsilon,\bar p_\epsilon\}=1,
 \label{eq:schutz-poisson-brackets}
\end{equation}
with vanishing crossed Poisson brackets. The complementary pair drops
out of the cosmological Hamiltonian, whereas the fluid contribution
becomes linear in the new momentum,

\begin{equation}
 \mathcal H_f=\frac{p_T}{a^{3\alpha}}.
 \label{eq:fluid-super-hamiltonian}
\end{equation}

The total super-Hamiltonian is consequently

\begin{equation}
 \mathcal H
 =-\frac{p_a^2}{24a}-6ka+\frac{p_T}{a^{3\alpha}}
 \approx0.
 \label{eq:original-super-hamiltonian}
\end{equation}

The clock interpretation is not based on linearity alone. The Hamilton
equations give

\begin{equation}
 \dot T=\frac{N}{a^{3\alpha}},
 \qquad
 \dot p_T=0.
 \label{eq:clock-hamilton-equations}
\end{equation}

For $a>0$ and a fixed lapse orientation $N>0$, $T$ is monotonic along
every regular classical trajectory. Its conjugate momentum is conserved.
Moreover, the chart used above has $p_T=e^s p_\epsilon^\gamma>0$, as
required by the positive-density matter sector. Combining
Eq.~\eqref{eq:pa-definition} with the constraint gives

\begin{equation}
 \left(\frac{\dot a}{Na}\right)^2+\frac{k}{a^2}
 =\frac{p_T}{6a^{3(1+\alpha)}}
 =\frac{\rho}{6},
 \label{eq:friedmann-equation}
\end{equation}
so that

\begin{equation}
 \rho(a)=\frac{p_T}{a^{3(1+\alpha)}}.
 \label{eq:fluid-density}
\end{equation}

Thus, $p_T$ fixes the conserved fluid content, while the sign of $\dot T$
is fixed by the lapse orientation on the regular cosmological domain.

For the comparison between reduced and extended descriptions, it is
useful to employ a deparametrized representative of the constraint. Since
$a^{3\alpha}$ is smooth, positive, and nonvanishing on the physical
domain $a>0$, the deparametrized representative is defined as

\begin{equation}
 \mathcal C:=a^{3\alpha}\mathcal H
 =p_T+H_\alpha(a,p_a)
 \approx0,
 \label{eq:deparametrized-constraint}
\end{equation}
where

\begin{equation}
 H_\alpha(a,p_a)
 =-\frac{1}{24}a^{3\alpha-1}p_a^2
  -6ka^{3\alpha+1}.
 \label{eq:reduced-hamiltonian}
\end{equation}

This rescaling must be distinguished from the canonical transformation
in Eqs.~\eqref{eq:schutz-clock-transformation} and
\eqref{eq:schutz-complementary-pair}. Multiplication by
$a^{3\alpha}$ preserves the classical constraint surface on $a>0$ and
amounts to the lapse redefinition $N_f=Na^{-3\alpha}$, but it is not a
canonical transformation. Classically equivalent representatives can
lead to inequivalent quantum realizations. Equation~\eqref{eq:deparametrized-constraint}
is selected because it is the representative
generated by the monotonic Schutz clock gauge: with $T=\tau$ its
multiplier is fixed to $N_f=1$, and solving it yields the same
$H_\alpha$ that generates reduced $T$-evolution. Using this common
representative in both descriptions isolates the question addressed
here---their compatibility once the clock, constraint representative,
and quantum realization have been fixed. Quantizing the unrescaled
super-Hamiltonian defines a different comparison and is not used as an
implicit benchmark.

\subsection{Relational reduction and quantum dynamics}
\label{subsec:relational-reduction}

The constraint in Eq.~\eqref{eq:deparametrized-constraint} is linear in
the clock momentum and can be solved as

\begin{equation}
 p_T=-H_\alpha(a,p_a).
 \label{eq:solve-clock-momentum}
\end{equation}

The formal reduced phase space is therefore the gravitational half-plane

\begin{equation}
 \Gamma_{\mathrm{red}}
 =\{(a,p_a)\,|\,a>0\},
 \qquad
 \omega_{\mathrm{red}}=da\wedge dp_a.
 \label{eq:reduced-phase-space}
\end{equation}

The positive-density choice $p_T>0$ further restricts the corresponding
classical sector to $H_\alpha<0$. This classical sign restriction will
later acquire a spectral, rather than pointwise, meaning in the quantum
theory. Starting from the constrained action
written with the rescaled lapse,

\begin{equation}
 S=\int d\tau\left[
 p_a\dot a+p_T\dot T
 -N_f\bigl(p_T+H_\alpha\bigr)
 \right],
 \label{eq:constrained-action}
\end{equation}
the gauge choice

\begin{equation}
 T=\tau,
 \qquad
 N_f=1,
 \label{eq:clock-gauge}
\end{equation}
gives the reduced action

\begin{equation}
 S_{\mathrm{red}}
 =\int dT\left[
 p_a\frac{da}{dT}-H_\alpha(a,p_a)
 \right].
 \label{eq:reduced-action}
\end{equation}

The physical Hamiltonian generating evolution with respect to the Schutz
clock is therefore $H_\alpha$. The reduced Hamilton equations are

\begin{align}
 \frac{da}{dT}
 &=-\frac{1}{12}a^{3\alpha-1}p_a,
 \label{eq:reduced-hamilton-a}\\
 \frac{dp_a}{dT}
 &=\frac{3\alpha-1}{24}a^{3\alpha-2}p_a^2
   +6k(3\alpha+1)a^{3\alpha}.
 \label{eq:reduced-hamilton-pa}
\end{align}

At the quantum level, the linear clock momentum converts the
Wheeler--DeWitt-type constraint into a relational Schr\"odinger equation.
With $\hat p_T=-i\hbar\partial_T$, the quantum constraint reads

\begin{equation}
 \left(-i\hbar\partial_T+\hat H_\alpha\right)\Psi(a,T)=0,
 \label{eq:quantum-constraint-wavefunction}
\end{equation}
or, equivalently,

\begin{equation}
 i\hbar\frac{\partial\Psi}{\partial T}
 =\hat H_\alpha\Psi.
 \label{eq:relational-schrodinger}
\end{equation}

Equation~\eqref{eq:reduced-hamiltonian} does not by itself determine a
quantum Hamiltonian on the half-line. To keep the ordering choice
explicit, consider the family

\begin{equation}
 \hat H_{\alpha,q}
 =\frac{\hbar^2}{24}a^{-q}\partial_a
 \left(a^{q+3\alpha-1}\partial_a\right)
 -6ka^{3\alpha+1},
 \label{eq:ordered-reduced-hamiltonian}
\end{equation}
which is formally symmetric in the kinematical space

\begin{equation}
 \mathcal H_q=L^2\!\left((0,\infty),a^q\,da\right).
 \label{eq:kinematical-hilbert-space}
\end{equation}

Formal symmetry is not sufficient to define the dynamics. A quantum
realization also requires an operator domain and, when necessary,
boundary conditions at $a=0$
\cite{ReedSimon1975,BonneauFarautValent2001,GitmanTyutinVoronov2012}.
Once a self-adjoint realization has been
fixed, the relational evolution is unitary and takes the form

\begin{equation}
 \Psi(T)=
 \exp\!\left[-\frac{i}{\hbar}(T-T_0)\hat H_{\alpha,q}\right]
 \Psi(T_0).
 \label{eq:unitary-relational-evolution}
\end{equation}

No classification of self-adjoint realizations is attempted here.
Instead, ordering, measure, domain, and boundary data are held fixed when
the reduced and extended descriptions are compared. The reconstruction
established below therefore concerns a specified quantum realization; it
does not assert an equivalence among all quantizations of classically
related constraints.

The reduced formulation provides a conventional relational dynamics,
but it does so only after solving the Hamiltonian constraint and removing
the clock pair $(T,p_T)$ from the phase space. The aim of the extended construction is to retain
that pair, impose the quantum constraint directly in the extended phase
space, and determine the conditions under which the reduced relational
state can be recovered from the constrained description. The
deformation-quantized phase-space language required for this construction
is introduced in Sec.~\ref{sec:deformation-constraints}.

\subsection{An explicit flat dust realization}
\label{subsec:flat-dust-example}

The admissible negative spectral sector, defined precisely in
Sec.~\ref{subsec:admissible-reduced-sector}, can be exhibited directly in a
simple cosmological case. Set $\alpha=0$, $k=0$, and choose $q=1/2$.
The reduced Hamiltonian and Hilbert space are

\begin{equation}
 \hat H_{0,1/2}
 =\frac{\hbar^2}{24}a^{-1/2}\partial_a
  \left(a^{-1/2}\partial_a\right),
 \qquad
 \mathcal H_{1/2}=L^2(\mathbb R_+,a^{1/2}da).
 \label{eq:dust-hamiltonian-a}
\end{equation}

Introduce

\begin{equation}
 x=\frac{4}{\sqrt3}a^{3/2},
 \qquad
 (V\psi)(x)=\frac{1}{\sqrt{2\sqrt3}}\,
 \psi\!\left[\left(\frac{\sqrt3 x}{4}\right)^{2/3}\right].
 \label{eq:dust-unitary-map}
\end{equation}

The map $V:\mathcal H_{1/2}\to L^2(\mathbb R_+,dx)$ is unitary and
transforms Eq.~\eqref{eq:dust-hamiltonian-a} into

\begin{equation}
 V\hat H_{0,1/2}V^{-1}=\frac{\hbar^2}{2}\frac{d^2}{dx^2}.
 \label{eq:dust-free-half-line-hamiltonian}
\end{equation}

For definiteness, take the Dirichlet realization
$D=H^2(\mathbb R_+)\cap H^1_0(\mathbb R_+)$. This operator is
self-adjoint and has purely absolutely continuous spectrum
$\sigma(\hat H_{0,1/2})=(-\infty,0]$. Its generalized eigenfunctions and
energies may be chosen as

\begin{equation}
 u_\kappa(x)=\sqrt{\frac{2}{\pi}}\sin(\kappa x),
 \qquad
 E_\kappa=-\frac{\hbar^2\kappa^2}{2},
 \qquad \kappa>0.
 \label{eq:dust-generalized-spectrum}
\end{equation}

Thus the spectral projector $\Pi=P((-\infty,0))$ onto the negative
spectral subspace (Eq.~\eqref{eq:negative-spectral-projector}) is the
identity up to the zero-energy spectral endpoint, which has zero spectral
measure. In this realization the positive-density Schutz sector is
nonempty and contains every normalizable reduced state.
For example, a normalized packet can be specified in the sine-transform
representation by

\begin{equation}
 g(\kappa)=\mathcal N
 \exp\!\left[-\frac{(\kappa-\kappa_0)^2}{4\sigma^2}
              -i\kappa x_0\right],
 \qquad \kappa>0,
 \label{eq:dust-gaussian-spectral-packet}
\end{equation}
where $\mathcal N$ normalizes the truncated Gaussian. The associated
pure-state density kernel is

\begin{equation}
 \rho_0(\kappa,\lambda)
 =g(\kappa)\overline{g(\lambda)},
 \qquad
 \hat\rho_0=\Pi\hat\rho_0\Pi,
 \label{eq:dust-admissibility}
\end{equation}
because $E_\kappa<0$ for every $\kappa>0$. Thus every such normalizable
packet lies in the admissible negative spectral sector. Its reconstruction
data on the extended phase space---the clock-momentum support and the
relational phase---are given in
Sec.~\ref{subsec:extended-reconstruction}, after the general
reconstruction map has been established.

This realization supplies an explicit boundary condition, continuous
spectral resolution, and a nonempty admissible sector. Other orderings,
curvatures, or boundary conditions can change the spectrum, which is why
the general construction keeps those data fixed.

\section{Deformation quantization and constraints}
\label{sec:deformation-constraints}

The reduced formulation of Sec.~\ref{sec:schutz} removes the clock
momentum by solving the Hamiltonian constraint before quantization. The
extended formulation retains instead the complete canonical chart

\begin{equation}
 \Gamma_{\mathrm{ext}}
 =\{(T,p_T;a,p_a)\,|\,a>0\},
 \qquad
 \omega_{\mathrm{ext}}=dT\wedge dp_T+da\wedge dp_a,
 \label{eq:extended-phase-space}
\end{equation}
and formulates the quantum constraint directly on the extended phase
space. If $\hat A$ is an operator in the fixed quantum realization
specified in Sec.~\ref{subsec:relational-reduction}, its Weyl symbol will
be denoted by $A_W$. Operator multiplication is represented by the star
product \cite{Folland1989,CurtrightFairlieZachos2014}

\begin{equation}
 (A_W\star B_W)(z)
 =A_W(z)
 \exp\!\left[
 \frac{i\hbar}{2}
 \overleftarrow{\partial}_{z^I}
 \Omega^{IJ}
 \overrightarrow{\partial}_{z^J}
 \right]
 B_W(z),
 \label{eq:extended-star-product}
\end{equation}
where $z=(T,p_T,a,p_a)$ and $\Omega^{IJ}$ is the inverse symplectic
matrix associated with Eq.~\eqref{eq:extended-phase-space}. In canonical
coordinates this becomes

\begin{align}
 A_W\star B_W
 =A_W\exp\!\Bigg\{\frac{i\hbar}{2}\Bigg[&
 \overleftarrow{\partial}_T\overrightarrow{\partial}_{p_T}
 -\overleftarrow{\partial}_{p_T}\overrightarrow{\partial}_T
 \nonumber\\
 &+\overleftarrow{\partial}_a\overrightarrow{\partial}_{p_a}
 -\overleftarrow{\partial}_{p_a}\overrightarrow{\partial}_a
 \Bigg]\Bigg\}B_W.
 \label{eq:canonical-extended-star-product}
\end{align}

The associated Moyal bracket is

\begin{equation}
 \{A_W,B_W\}_{M}
 :=\frac{1}{i\hbar}
 \left(A_W\star B_W-B_W\star A_W\right),
 \label{eq:moyal-bracket}
\end{equation}
and reduces to the Poisson bracket at leading order in $\hbar$.

For a density operator $\hat\rho$ associated with one canonical
pair, the normalized Wigner convention adopted here is

\begin{equation}
 W=\frac{1}{2\pi\hbar}\rho_W.
 \label{eq:wigner-normalization}
\end{equation}

For the extended system, which contains two canonical pairs,
the corresponding convention is
$W_{\mathrm{ext}}=(2\pi\hbar)^{-2}
(\rho_{\mathrm{ext}})_W$, understood distributionally.
This normalization convention does not imply an ordinary
kinematical trace normalization of the extended state.

Hermiticity of $\hat\rho$ implies that $W$ is real, whereas positivity
remains an operator property and is not equivalent to pointwise
positivity of $W$ \cite{Wigner1932,Hillery1984,Hudson1974}.

For a density operator on $L^2(\mathbb R,da)$ with integral kernel
$\widetilde\rho(a_+,a_-)$, this convention corresponds to the standard
Weyl--Wigner transform
\begin{equation}
 W(a,p_a)
 =\frac{1}{2\pi\hbar}
 \int_{-\infty}^{\infty}d\xi\,
 e^{-ip_a\xi/\hbar}\,
 \widetilde\rho\!\left(
 a+\frac{\xi}{2},
 a-\frac{\xi}{2}
 \right),
 \label{eq:full-line-wigner-transform}
\end{equation}
where $a=(a_++a_-)/2$ is the midpoint coordinate and
$\xi=a_+-a_-$ is the relative coordinate
\cite{Hillery1984,Folland1989}.

The gravitational configuration variable requires an additional step
because $a$ belongs to the half-line and the reduced Hilbert space carries
the measure $a^q da$. This weight is first removed with the unitary map

\begin{equation}
 U_q:\mathcal H_q\longrightarrow L^2(\mathbb R_+,da),
 \qquad
 (U_q\psi)(a)=a^{q/2}\psi(a),
 \label{eq:measure-removing-unitary}
\end{equation}
after which the zero-extension isometry is applied,

\begin{equation}
 J:L^2(\mathbb R_+,da)\longrightarrow L^2(\mathbb R,da),
 \qquad
 (J\phi)(a)=\Theta(a)\phi(a),
 \label{eq:zero-extension-map}
\end{equation}
where $\Theta(a)$ denotes the Heaviside step function, taken here as
$\Theta(a)=1$ for $a>0$ and $\Theta(a)=0$ for $a<0$; its value at
$a=0$ is immaterial for the present $L^2$ construction.

For an operator $\hat A$ on $\mathcal H_q$, its operational full-line
representative is

\begin{equation}
 \widetilde A
 :=J U_q\hat A U_q^{-1}J^\dagger.
 \label{eq:operational-full-line-operator}
\end{equation}

Since $J^\dagger J=I$ on $L^2(\mathbb R_+,da)$, this embedding preserves
operator composition on its natural range,

\begin{equation}
 \widetilde{AB}=\widetilde A\,\widetilde B.
 \label{eq:embedded-composition}
\end{equation}
The embedding does not, however, preserve the identity: $JJ^\dagger=P_+$,
where $P_+$ denotes multiplication by $\Theta(a)$ on $L^2(\mathbb R,da)$,
so that the half-line identity is represented by $\widetilde I=P_+$ rather
than by the full-line identity.

Accordingly, $A_W$ is defined as the ordinary full-line Weyl symbol of
$\widetilde A$. Consequently, the Weyl correspondence
turns Eq.~\eqref{eq:embedded-composition} into the star product in
Eq.~\eqref{eq:canonical-extended-star-product}. For a density operator
with integral kernel $\rho_q(a_+,a_-)$ relative to the measure $a^q da$,
the resulting gravitational Wigner function is, for $a>0$,

\begin{align}
 W_\rho^{(0)}(a,p_a)
 =\frac{1}{2\pi\hbar}
 \int_{-2a}^{2a}d\xi\,
 e^{-ip_a\xi/\hbar}
 &\left(a+\frac{\xi}{2}\right)^{q/2}
 \left(a-\frac{\xi}{2}\right)^{q/2}
 \nonumber\\[-2mm]
 &\times
 \rho_q\!\left(a+\frac{\xi}{2},a-\frac{\xi}{2}\right).
 \label{eq:half-line-operational-wigner}
\end{align}

It vanishes for $a<0$. The restricted integration range is not an
additional boundary condition; it follows from the support of the
zero-extended kernel.

This construction defines the phase-space representation used in the
reconstruction theorem, but it is not an intrinsic quantization of the
half-line. The self-adjoint domain and boundary condition at $a=0$ remain
independent data of the operator realization
\cite{BonneauFarautValent2001,GitmanTyutinVoronov2012}. They enter the
phase-space state through the spectral eigenfunctions occurring in its
kernel. The differential Moyal equations are therefore understood on the
interior $a>0$; possible boundary contributions belong to the
distributional full-line symbols and are not replaced by the local
bidifferential formula alone.

Finally, because the clock runs over a noncompact domain and the
constraint fixes its conjugate momentum, physical extended states may
also be distributional in the clock sector. None of the algebraic
manipulations below relies on pointwise positivity or ordinary trace
normalization in that sector.

For the representative fixed in Eq.~\eqref{eq:deparametrized-constraint},
the constraint operator is

\begin{equation}
 \hat{\mathcal C}
 =\hat p_T+\hat H_{\alpha,q},
 \label{eq:constraint-operator}
\end{equation}
and its Weyl symbol has the form

\begin{equation}
 \mathcal C_W=p_T+H_W(a,p_a).
 \label{eq:constraint-weyl-symbol}
\end{equation}

Here $H_W$ denotes the operational Weyl symbol, in the sense of
Eq.~\eqref{eq:operational-full-line-operator}, of the chosen self-adjoint
realization of $\hat H_{\alpha,q}$. It must not be identified
automatically with the classical function $H_\alpha$: factor ordering
may generate $\hbar$-dependent contributions to the symbol. Keeping the
notation $H_W$ explicit prevents the comparison between reduced and
extended formulations from silently changing the quantum realization.

The full-line representative of the constraint requires one further
specification. Under the embedding, the clock term $\hat p_T\otimes I$ is
mapped to $\hat p_T\otimes P_+$, whose Weyl symbol is $p_T\,\Theta(a)$,
whereas $\widetilde H$ vanishes on the orthogonal complement of
$\operatorname{Ran}P_+$. All extended states considered below lie in the
embedded range, in the sense that $(I\otimes P_+)\widetilde\rho
=\widetilde\rho=\widetilde\rho\,(I\otimes P_+)$, and on this range
$\hat p_T\otimes P_+$ and $\hat p_T\otimes I$ act identically from either
side. Equation~\eqref{eq:constraint-weyl-symbol} is therefore adopted as
the full-line representative of the constraint: its left and right star
actions on embedded states coincide with those of $p_T\,\Theta(a)+H_W$.
Throughout, star products involving distributional symbols are understood
as Weyl images of the corresponding operator compositions on the embedded
range, not as consequences of the formal bidifferential expansion alone.

At the operator level, a density operator supported on the physical
subspace is annihilated by the constraint from both sides,

\begin{equation}
 \hat{\mathcal C}\hat\rho=0,
 \qquad
 \hat\rho\hat{\mathcal C}=0.
 \label{eq:bilateral-operator-constraint}
\end{equation}

Under the Weyl map, these conditions become

\begin{equation}
 \mathcal C_W\star W=0,
 \qquad
 W\star\mathcal C_W=0.
 \label{eq:bilateral-star-constraint}
\end{equation}

The two equations carry complementary information. Their antisymmetric
and symmetric combinations encode, respectively, transport in the
internal time and support on the quantum constraint surface. Because the
constraint is linear in $p_T$, the left and right clock-momentum actions
truncate exactly at first order:

\begin{align}
 p_T\star W
 &=p_TW-\frac{i\hbar}{2}\partial_TW,
 \label{eq:left-clock-star-action}\\
 W\star p_T
 &=p_TW+\frac{i\hbar}{2}\partial_TW.
 \label{eq:right-clock-star-action}
\end{align}

Subtracting the two equations in
Eq.~\eqref{eq:bilateral-star-constraint} therefore gives

\begin{equation}
 -i\hbar\partial_TW
 +H_W\star W-W\star H_W=0,
 \label{eq:antisymmetric-constraint}
\end{equation}
or equivalently

\begin{equation}
 \boxed{
 \partial_TW=\{H_W,W\}_{M}.
 }
 \label{eq:extended-moyal-evolution}
\end{equation}

The antisymmetric part of the bilateral constraint is therefore the
exact Moyal evolution with respect to the Schutz clock. In the classical
limit it reduces to Hamiltonian transport generated by $H_\alpha$,
provided $H_W=H_\alpha+O(\hbar)$.

Adding the same two equations eliminates the clock derivative and yields

\begin{equation}
 \boxed{
 p_TW
 +\frac{1}{2}
 \left(H_W\star W+W\star H_W\right)=0.
 }
 \label{eq:quantum-shell-condition}
\end{equation}

This symmetric equation is the quantum counterpart of the classical
shell condition $p_T+H_\alpha=0$. It does not generate evolution;
instead, it restricts the dependence of the extended state on $p_T$ and
retains information that is absent after classical reduction. Except in
special commuting sectors, it cannot be replaced by the pointwise
relation $p_T=-H_W$, since doing so would discard the symmetric
left--right star action that the reconstruction must preserve.

Equations~\eqref{eq:extended-moyal-evolution} and
\eqref{eq:quantum-shell-condition} are jointly equivalent to the
bilateral system in Eq.~\eqref{eq:bilateral-star-constraint}. The first
equation reproduces the relational dynamics expected from the reduced
description, whereas the second determines whether a phase-space state
belongs to the quantum constraint surface. This separation is the
structural reason for retaining both star equations.

Let $\hat\rho_{\mathrm{red}}(T)$ evolve with the fixed Hamiltonian
$\hat H_{\alpha,q}$ of Sec.~\ref{subsec:relational-reduction}. Its Weyl
symbol $W_{\mathrm{red}}(a,p_a;T)$ satisfies

\begin{equation}
 \partial_TW_{\mathrm{red}}
 =\{H_W,W_{\mathrm{red}}\}_{M},
 \label{eq:reduced-moyal-evolution}
\end{equation}
but carries no $p_T$ dependence and therefore does not by itself satisfy
the symmetric shell condition. Consequently, solving the evolution
equation is necessary but not sufficient for constructing a physical
state on $\Gamma_{\mathrm{ext}}$. The remaining problem is to determine
which reduced states admit an extension in the clock momentum and how
that extension can satisfy both sides of the quantum constraint. This
defines the reconstruction problem addressed in the following section.

\section{Reduced and extended formulations}
\label{sec:reconstruction}

\subsection{The admissible reduced sector}
\label{subsec:admissible-reduced-sector}

Let $\hat H\equiv\hat H_{\alpha,q}$ denote the self-adjoint reduced
Hamiltonian fixed in Sec.~\ref{subsec:relational-reduction}, and let
$\hat\rho_0$ be a positive trace-class operator on $\mathcal H_q$. The
spectral resolution of $\hat H$ will be written as

\begin{equation}
 \hat H
 =\int_{\sigma(\hat H)} E\,P(dE),
 \label{eq:spectral-resolution}
\end{equation}
where $P(dE)$ is the projection-valued spectral measure. The relational
evolution of a reduced density operator is

\begin{equation}
 \hat\rho_{\mathrm{red}}(T)
 =e^{-\frac{i}{\hbar}(T-T_0)\hat H}
 \hat\rho_0
 e^{\frac{i}{\hbar}(T-T_0)\hat H}.
 \label{eq:reduced-density-evolution}
\end{equation}

The positive-density Schutz sector imposes $p_T>0$. Since the classical
constraint gives $p_T=-H_\alpha$, its quantum counterpart selects the
negative spectral subspace of the fixed Hamiltonian. The negative
spectral projector and its range are therefore defined as

\begin{equation}
 \Pi:=P(( -\infty,0)),
 \qquad
 \mathcal K:=\operatorname{Ran}\Pi,
 \label{eq:negative-spectral-projector}
\end{equation}
and a reduced state is called admissible when

\begin{equation}
 \boxed{
 \hat\rho_0=\Pi\hat\rho_0\Pi.
 }
 \label{eq:admissibility-condition}
\end{equation}

Because $\Pi$ is a spectral projector of $\hat H$, this condition is
preserved by the relational evolution. It is the precise quantum version
of the classical restriction $H_\alpha<0$ identified in
Sec.~\ref{subsec:relational-reduction}. No pointwise sign condition on a
Wigner function is involved.

To display the reconstruction explicitly, it is convenient to use a
generalized spectral
basis $|x\rangle=|E_x,\lambda_x\rangle$ in the rigged-Hilbert-space
sense \cite{deLaMadrid2005}, where $\lambda_x$ denotes any
degeneracy labels and integration with $d\nu(x)$ includes the appropriate
discrete sums and continuous spectral measures. On the admissible
subspace, $E_x<0$, and

\begin{equation}
 \hat\rho_0
 =\int_- d\nu(x)\int_-d\nu(y)\,
 \rho_0(x,y)|x\rangle\langle y|.
 \label{eq:initial-density-spectral-kernel}
\end{equation}

The subscript on the integrals indicates restriction to the negative
spectrum. For continuous spectral components, the dyads and their
coefficients are understood in the usual generalized-eigenvector sense.
Equation~\eqref{eq:reduced-density-evolution} then gives

\begin{equation}
 \hat\rho_{\mathrm{red}}(T)
 =\int_- d\nu(x)\int_-d\nu(y)\,
 \rho_0(x,y)
 e^{-\frac{i}{\hbar}(E_x-E_y)(T-T_0)}
 |x\rangle\langle y|.
 \label{eq:reduced-density-spectral-evolution}
\end{equation}

Let

\begin{equation}
 \Omega_{xy}(a,p_a)
 :=\frac{1}{2\pi\hbar}\,
 \mathcal W_a\!\left[|x\rangle\langle y|\right]
 \label{eq:cross-weyl-symbol}
\end{equation}
denote the normalized gravitational Wigner transform of the spectral dyad,
where $\mathcal W_a$ is the operational Weyl symbol map defined through
Eq.~\eqref{eq:operational-full-line-operator}. The prefactor is the one in
Eq.~\eqref{eq:wigner-normalization}, so that
Eq.~\eqref{eq:reduced-wigner-spectral-form} below is the normalized Wigner
function of $\hat\rho_{\mathrm{red}}(T)$. Since the Weyl map converts
operator multiplication into the star product and the prefactor is
constant, these cross symbols obey the exact left and right star-eigenvalue equations

\begin{equation}
 H_W\star_a\Omega_{xy}=E_x\Omega_{xy},
 \qquad
 \Omega_{xy}\star_a H_W=E_y\Omega_{xy},
 \label{eq:cross-star-eigenvalue-equations}
\end{equation}
where $\star_a$ denotes the gravitational part of the product in
Eq.~\eqref{eq:canonical-extended-star-product}. The reduced Wigner state
is consequently

\begin{equation}
 W_{\mathrm{red}}(a,p_a;T)
 =\int_- d\nu(x)\int_-d\nu(y)\,
 \rho_0(x,y)
 e^{-\frac{i}{\hbar}(E_x-E_y)(T-T_0)}
 \Omega_{xy}(a,p_a).
 \label{eq:reduced-wigner-spectral-form}
\end{equation}

Off-diagonal terms are essential in this expression. Restricting the
construction to $x=y$ would retain stationary mixtures but would remove
the coherences responsible for nontrivial relational evolution.

\subsection{Reconstruction on the extended phase space}
\label{subsec:extended-reconstruction}

The clock dependence required by the bilateral constraint can be read
from the spectral dyads. For each pair $(x,y)$, define the distribution

\begin{equation}
 \chi_{xy}(T,p_T)
 :=e^{-\frac{i}{\hbar}(E_x-E_y)(T-T_0)}
 \delta\!\left(p_T+\frac{E_x+E_y}{2}\right).
 \label{eq:clock-reconstruction-kernel}
\end{equation}

Its normalization is fixed by the requirement that
$\int dp_T\,\chi_{xy}$ reproduce the phase factor in
Eq.~\eqref{eq:reduced-wigner-spectral-form}. With generalized
clock-momentum eigenstates normalized as
$\langle T|p_T\rangle=(2\pi\hbar)^{-1/2}e^{\frac{i}{\hbar}p_TT}$, the
normalized Wigner transform of the dyad
$|p_T=-E_x\rangle\langle p_T=-E_y|$ is

\begin{equation}
 \frac{1}{2\pi\hbar}\,
 e^{-\frac{i}{\hbar}(E_x-E_y)T}\,
 \delta\!\left(p_T+\frac{E_x+E_y}{2}\right),
 \label{eq:clock-dyad-wigner}
\end{equation}
so $\chi_{xy}$ differs from it only by the constant factor
$2\pi\hbar\,e^{\frac{i}{\hbar}(E_x-E_y)T_0}$. This factor is exactly the
one carried by the generalized embedding $\mathcal V$ constructed below,
for which
$(\mathcal V|x\rangle)(T)=e^{-\frac{i}{\hbar}E_x(T-T_0)}|x\rangle
=(2\pi\hbar)^{1/2}e^{\frac{i}{\hbar}E_xT_0}\,
\langle T|p_T=-E_x\rangle\,|x\rangle$. Consequently, the extended state defined below is the normalized
two-pair Wigner transform of the formal operator
$\mathcal V\hat\rho_0\mathcal V^\dagger$.
The factor $2\pi\hbar$ compensates the Fourier normalization
of the generalized clock-momentum eigenstates and ensures
the stated clock-momentum marginal.
Physical normalization is specified by the induced inner
product in Eq.~\eqref{eq:induced-physical-inner-product};
no finite kinematical trace or explicit division by a gauge
volume is asserted.
The reconstructed extended state is then

\begin{equation}
 \boxed{
 W_{\mathrm{ext}}(T,p_T,a,p_a)
 =\int_- d\nu(x)\int_-d\nu(y)\,
 \rho_0(x,y)\,
 \chi_{xy}(T,p_T)\,
 \Omega_{xy}(a,p_a).
 }
 \label{eq:extended-reconstruction-map}
\end{equation}

The flat-dust model of Sec.~\ref{subsec:flat-dust-example} can now be
completed without anticipating the general construction. Setting
$x=\kappa$ and $y=\lambda$ in
Eq.~\eqref{eq:extended-reconstruction-map}, and using
$E_\kappa=-\hbar^2\kappa^2/2$ and
$E_\lambda=-\hbar^2\lambda^2/2$, gives

\begin{equation}
 p_T=-\frac{E_\kappa+E_\lambda}{2}
 =\frac{\hbar^2}{4}(\kappa^2+\lambda^2)>0,
 \qquad
 e^{-\frac{i}{\hbar}(E_\kappa-E_\lambda)(T-T_0)}
 =e^{\frac{i\hbar}{2}(\kappa^2-\lambda^2)(T-T_0)}.
 \label{eq:dust-reconstruction-data}
\end{equation}

Thus the example prepared in Sec.~\ref{subsec:flat-dust-example}
realizes explicitly both pieces of the spectral reconstruction: the
average of the two energies fixes the positive clock-momentum support,
whereas their difference fixes the off-diagonal relational phase.

This expression has a direct operator interpretation. Up to the constant
factor discussed after Eq.~\eqref{eq:clock-dyad-wigner}, the clock factor
is the Wigner transform of the dyad formed by generalized momentum
eigenstates with

\begin{equation}
 p_T=-E_x
 \quad\hbox{on the ket},
 \qquad
 p_T=-E_y
 \quad\hbox{on the bra}.
 \label{eq:clock-momentum-matching}
\end{equation}

The Wigner transform associates such an off-diagonal dyad with the
average momentum $-(E_x+E_y)/2$, while its oscillatory $T$ dependence is
controlled by the energy difference $E_x-E_y$. The delta distribution
in Eq.~\eqref{eq:clock-reconstruction-kernel} therefore encodes both
left and right constraint data; it is not obtained by inserting a single
classical energy into $p_T=-H_\alpha$.

The operator meaning of this expression is formulated on a Gelfand
triple, because clock-momentum eigenstates belong to the dual test space
instead of the kinematical Hilbert space \cite{deLaMadrid2005}. Let

\begin{equation}
 \Phi
 =\mathcal S(\mathbb R_T)\widehat\otimes\mathcal D_-
 \label{eq:extended-test-space}
\end{equation}
be a test space, where $\mathcal D_-$ is a dense invariant domain in
$\mathcal K$, and define

\begin{equation}
 \mathcal R F
 :=\Pi\int_{-\infty}^{\infty}dT\,
 e^{\frac{i}{\hbar}(T-T_0)\hat H}F(T),
 \qquad F\in\Phi.
 \label{eq:reduction-map-test-space}
\end{equation}

The generalized embedding $\mathcal V:\mathcal K\to\Phi'$ is defined
by duality through

\begin{equation}
 \langle\mathcal V\psi,F\rangle
 :=\langle\psi,\mathcal R F\rangle_{\mathcal K}.
 \label{eq:generalized-physical-embedding}
\end{equation}

With the clock-momentum normalization specified above, the
spectral representation of this embedding is

\begin{equation}
 \mathcal V|x\rangle
 =\sqrt{2\pi\hbar}\,
 e^{\frac{i}{\hbar}E_xT_0}
 |p_T=-E_x\rangle\otimes|x\rangle.
 \label{eq:spectral-generalized-embedding}
\end{equation}

The notation

\begin{equation}
 \hat\rho_{\mathrm{ext}}
 =\mathcal V\hat\rho_0\mathcal V^\dagger
 \label{eq:extended-density-operator}
\end{equation}
is therefore shorthand for the sesquilinear distribution

\begin{equation}
 \rho_{\mathrm{ext}}[F,G]
 :=\langle\mathcal R F|\hat\rho_0|\mathcal R G\rangle_{\mathcal K},
 \qquad F,G\in\Phi.
 \label{eq:positive-distributional-density}
\end{equation}

This definition makes positivity precise:

\begin{equation}
 \rho_{\mathrm{ext}}[F,F]\geq0
 \qquad\text{for every }F\in\Phi.
 \label{eq:distributional-positivity}
\end{equation}

It also preserves hermiticity, and
$\rho_0(y,x)=\rho_0(x,y)^*$ ensures that the extended Weyl symbol is real.
If $\operatorname{Tr}_{\mathcal K}\hat\rho_0=1$, the induced physical
normalization is fixed by

\begin{equation}
 \langle\mathcal V\psi,\mathcal V\varphi\rangle_{\mathrm{phys}}
 :=\langle\psi,\varphi\rangle_{\mathcal K}.
 \label{eq:induced-physical-inner-product}
\end{equation}

No ordinary trace normalization of
$\hat\rho_{\mathrm{ext}}$ in the kinematical clock Hilbert space is
claimed.

The constraint is satisfied weakly as well. Integration by parts in
Eq.~\eqref{eq:reduction-map-test-space}, with the rapid decay of the test
functions, gives

\begin{equation}
 \mathcal R\hat{\mathcal C}F=0.
 \label{eq:weak-constraint-reduction-map}
\end{equation}

Hence

\begin{equation}
 \rho_{\mathrm{ext}}[\hat{\mathcal C}F,G]=0,
 \qquad
 \rho_{\mathrm{ext}}[F,\hat{\mathcal C}G]=0,
 \label{eq:weak-bilateral-constraint}
\end{equation}
which is the distributional operator statement underlying the two star
constraints proved below.

Since $E_x<0$ and $E_y<0$ throughout the admissible sector,

\begin{equation}
 -\frac{E_x+E_y}{2}>0.
 \label{eq:positive-clock-support}
\end{equation}

The reconstructed state is consequently supported entirely at positive
values of the Wigner clock momentum. More strongly, the two-sided
condition in Eq.~\eqref{eq:admissibility-condition} ensures that the ket
and bra clock momenta, $-E_x$ and $-E_y$, are separately positive.
Positivity of the average momentum alone would not be sufficient: a
mixed-sign pair of energies can have
$-(E_x+E_y)/2>0$ even though one side of the corresponding operator dyad
lies outside the positive clock-momentum sector.

The reduced state is recovered as the $p_T$ marginal of the extended
state:

\begin{equation}
 \boxed{
 \int_{-\infty}^{\infty}dp_T\,
 W_{\mathrm{ext}}(T,p_T,a,p_a)
 =W_{\mathrm{red}}(a,p_a;T).
 }
 \label{eq:reduction-as-clock-momentum-marginal}
\end{equation}

Thus the reconstruction map is injective on the admissible spectral
sector: if two reconstructed states coincide, their $p_T$ marginals
coincide, and since the Weyl--Wigner correspondence is injective, so do
the corresponding reduced density operators and their initial data
$\hat\rho_0$. Equation~\eqref{eq:reduction-as-clock-momentum-marginal}
also shows that no decoherence or diagonal-energy approximation is
involved in passing back to the reduced description.

\subsection{Bilateral constraint and reconstruction theorem}
\label{subsec:reconstruction-proof}

It remains to verify that Eq.~\eqref{eq:extended-reconstruction-map}
satisfies both star constraints. For the clock kernel in
Eq.~\eqref{eq:clock-reconstruction-kernel},

\begin{equation}
 \partial_T\chi_{xy}
 =-\frac{i}{\hbar}(E_x-E_y)\chi_{xy}.
 \label{eq:clock-kernel-time-derivative}
\end{equation}

Let $\star_T$ denote the clock part of the extended star product. Using
Eqs.~\eqref{eq:left-clock-star-action} and
\eqref{eq:right-clock-star-action}, one obtains

\begin{align}
 p_T\star_T\chi_{xy}
 &=\left[p_T-\frac{E_x-E_y}{2}\right]\chi_{xy}
 =-E_x\chi_{xy},
 \label{eq:left-clock-kernel-eigenvalue}\\
 \chi_{xy}\star_T p_T
 &=\left[p_T+\frac{E_x-E_y}{2}\right]\chi_{xy}
 =-E_y\chi_{xy},
 \label{eq:right-clock-kernel-eigenvalue}
\end{align}
where the final equalities hold distributionally on the support
$p_T=-(E_x+E_y)/2$. Combining these relations with
Eq.~\eqref{eq:cross-star-eigenvalue-equations} gives, term by term,

\begin{align}
 (p_T+H_W)\star
 \bigl(\chi_{xy}\Omega_{xy}\bigr)
 &=(-E_x+E_x)\chi_{xy}\Omega_{xy}=0,
 \label{eq:left-reconstructed-constraint}\\
 \bigl(\chi_{xy}\Omega_{xy}\bigr)\star
 (p_T+H_W)
 &=(-E_y+E_y)\chi_{xy}\Omega_{xy}=0.
 \label{eq:right-reconstructed-constraint}
\end{align}

Linearity then proves

\begin{equation}
 \boxed{
 \mathcal C_W\star W_{\mathrm{ext}}=0,
 \qquad
 W_{\mathrm{ext}}\star\mathcal C_W=0.
 }
 \label{eq:reconstructed-bilateral-constraint}
\end{equation}

The result may be summarized as follows. Given a self-adjoint realization
of $\hat H$, every reduced density operator supported on
$\mathcal K$ admits the distributional extension in
Eq.~\eqref{eq:extended-reconstruction-map}. The extension is supported
on $p_T>0$, defines the positive sesquilinear distribution in
Eq.~\eqref{eq:positive-distributional-density}, satisfies the bilateral
quantum constraint, evolves with the same relational phases as the
reduced state, and returns the reduced Wigner function under the marginal in
Eq.~\eqref{eq:reduction-as-clock-momentum-marginal}. Within the spectral
class constructed above, the marginal is a left inverse of the
reconstruction map and therefore determines the extension uniquely.

This statement establishes compatibility, not an unrestricted
commutation of reduction and quantization. It depends on the chosen
constraint representative, the fixed operator realization, and the
negative spectral support of the reduced state. Nor does it claim that
every distributional solution of the bilateral equations arises from
the reconstruction map. The result identifies a controlled physical
sector in which the reduced relational description is embedded in the
extended constrained theory without losing either quantum coherences or
the symmetric shell condition.

\section{Discussion}
\label{sec:discussion}

The reconstruction established in Sec.~\ref{sec:reconstruction} gives a
precise relation between two descriptions of Schutz quantum cosmology.
In the reduced formulation, the fluid
variable $T$ is chosen as an internal time and the Hamiltonian constraint
is solved before quantization. The resulting theory has an ordinary
relational evolution generated by $\hat H$. In the extended formulation,
the clock pair is retained and physical states are selected by the
bilateral quantum constraint. The reconstruction shows that these
descriptions are compatible on a precisely identified spectral sector,
without requiring the Hamiltonian constraint to be discarded at the
quantum level.

The bilateral character of the extended constraint is essential to this
statement. Its antisymmetric part reproduces the Moyal evolution of the
reduced state, but that equation alone only determines transport with
respect to $T$. The symmetric part supplies the missing shell condition
and fixes the clock-momentum dependence of the extended state. Thus the
reduced theory retains the relational dynamics but no longer displays
the full quantum support condition. Reconstruction restores this
information by correlating the gravitational spectral data with the
clock momentum.

For diagonal spectral components, this correlation takes the familiar
form $p_T=-E$. Off the diagonal, however, the Wigner clock momentum is
supported at

\begin{equation}
 p_T=-\frac{E_x+E_y}{2},
 \label{eq:discussion-average-energy}
\end{equation}
while the $T$ dependence is governed by the difference $E_x-E_y$. The
average and difference of the two spectral values therefore encode
distinct aspects of the extended state: the former locates the quantum
shell in clock momentum, and the latter generates relational phase
evolution. This distinction explains why a reconstruction based only on
diagonal energy distributions would be incomplete. Such a restriction
would reproduce stationary mixtures but eliminate the coherences needed
for general quantum evolution.

The physical sign of the Schutz momentum also becomes sharper in the
spectral description. The positive-density chart has $p_T>0$, and the
operator constraint pairs it with the negative spectrum of $\hat H$.
Accordingly, $\hat\rho_0=\Pi\hat\rho_0\Pi$ implements the classically
selected matter sector on both the ket and bra. Positivity of only the
average Wigner momentum in Eq.~\eqref{eq:discussion-average-energy}
would allow one spectral leg to lie outside that sector. The flat dust
realization in Sec.~\ref{subsec:flat-dust-example} shows explicitly that
the admissible space need not be empty: there $\Pi$ covers the complete
normalizable reduced Hilbert space.

Several choices have been held fixed from the outset, so the result
makes no claim to establish commutativity of reduction and quantization
in general. First, the
constraint quantized in the extended theory is the
deparametrized representative
$\mathcal C=a^{3\alpha}\mathcal H=p_T+H_\alpha$. Although this
representative has the same classical constraint surface as
$\mathcal H$ on $a>0$, rescaling changes the object to be quantized. The
Schutz gauge selects this representative because it generates the
reduced $T$-evolution. The comparison therefore uses it consistently in
the reduced and extended theories. A theory obtained by quantizing the
unrescaled super-Hamiltonian remains outside the comparison.

Second, the ordering, measure, domain, boundary conditions, and
self-adjoint realization of the reduced Hamiltonian are part of the
definition of the model. These data determine both the spectrum used in
the admissibility condition and the Weyl symbol $H_W$ entering the star
constraints. Changing them may alter the negative spectral subspace and
hence the set of reconstructible states. If the chosen realization has
no negative spectrum, the reconstructed positive-density sector is
empty; zero modes correspond to $p_T=0$ and are excluded by the strict
orientation adopted here. The use of a common fixed realization in both
descriptions is therefore a substantive consistency condition rather
than a notational convenience.

This qualification is particularly important because the scale factor
lives on the half-line. The canonical Moyal product provides the local
symbol calculus used in the reconstruction, but it does not by itself
encode the global behavior at $a=0$. Boundary conditions and operator
domains remain indispensable. A formulation based on a different global
phase-space quantization, such as an affine construction adapted to the
half-line \cite{Gouba2021}, could modify the symbol calculus while preserving the basic
logic of a bilateral constraint. Establishing the precise relation
between such formulations lies beyond the present work.

The clock sector introduces a different analytic qualification. Exact
solutions of the constraint correlate gravitational energies with
generalized eigenstates of $\hat p_T$ and are consequently
distributional on the noncompact clock phase space. This is the
phase-space counterpart of the familiar fact that solutions of a
Hamiltonian constraint are generally not normalizable in the
kinematical inner product before the gauge volume is removed. The
sesquilinear formulation in
Eq.~\eqref{eq:positive-distributional-density} defines a positive
distribution and induces the physical inner product in
Eq.~\eqref{eq:induced-physical-inner-product}; the extended state remains
non-trace-class in the kinematical clock sector. In refined algebraic
quantization, a rigging map and physical product can instead be obtained
by averaging the one-parameter group generated by the constraint
\cite{Marolf1995RAQ,Giulini2000GroupAveraging}. A phase-space counterpart,
in which physical Wigner distributions are obtained by star-exponentiating
and averaging the constraints, was formulated in
Ref.~\cite{BerraMontielMolgado2020}. The map $\mathcal V$ has
the same formal target---distributional solutions of the constraint---but
here it is fixed by spectral matching rather than derived from a group
average. Equality with either group-averaged construction would require a specified
averaging domain and normalization and is not assumed here.

The Weyl representation also carries relational Dirac observables. For
an operator $\hat A$ on the reduced gravitational Hilbert space, define

\begin{equation}
 \hat O_A(\tau)
 =e^{-\frac{i}{\hbar}(T-\tau)\hat H}\,
  \hat A\,
  e^{\frac{i}{\hbar}(T-\tau)\hat H}.
 \label{eq:relational-dirac-observable}
\end{equation}

It obeys $[\hat{\mathcal C},\hat O_A(\tau)]=0$ and represents the value of
$A$ when the Schutz clock reads $\tau$, in the standard complete-observable
sense \cite{Dittrich2006}. Its Weyl symbol consequently satisfies
$\mathcal C_W\star O_{A,W}-O_{A,W}\star\mathcal C_W=0$. Because the Weyl
map preserves operator multiplication, commutators and the algebra of
these relational observables are represented by star commutators in the
extended phase space.

Within these limits, the mechanism does not depend on the detailed
functional form of the FLRW Hamiltonian. Its essential ingredients are a
deparametrized constraint linear in a clock momentum,

\begin{equation}
 \mathcal C=p_T+H,
 \label{eq:generic-deparametrized-constraint}
\end{equation}
a self-adjoint realization of $\hat H$, and a spectral sector compatible
with the physical orientation of the clock. Whenever these conditions
hold, spectral dyads of $\hat H$ can be paired with clock-momentum dyads
in the same way as in Eq.~\eqref{eq:spectral-generalized-embedding}. The
average-energy shell and difference-energy evolution then follow from
the left and right actions of the constraint. This suggests that the
construction may extend to other deparametrizable minisuperspace models
and to matter clocks different from the Schutz fluid.

Any such extension remains sensitive to the choice of clock. A different
internal time generally leads to a different reduced
Hamiltonian, spectral decomposition, and notion of positive frequency.
The multiple-choice aspect of the problem of time remains open. The
present construction provides a controlled test of a chosen relational
clock against an extended constrained quantization. Comparing the reconstructed sectors associated with
different clocks would be a natural continuation of this analysis.

Further applications may examine other equations of state and spatial
curvatures, the semiclassical behavior of reconstructed states, and more
general relational observables. These questions require concrete spectral information about
$\hat H_{\alpha,q}$ and its self-adjoint realizations. They do not modify
the reconstruction mechanism itself, but they are necessary for
determining how its admissible sector is realized in specific
cosmological models.

\section{Conclusion}
\label{sec:conclusion}

This work has identified sufficient conditions under which reduced Schutz-clock
dynamics embeds consistently into an extended deformation-quantized
description. The construction uses the same deparametrized representative
$\mathcal C=p_T+H$ and the same self-adjoint realization of $\hat H$ in
both formulations.

In deformation quantization, the extended constraint must act on the
phase-space state from both sides. The antisymmetric part
of this bilateral system was shown to give the exact Moyal evolution with respect to
the Schutz clock, while the symmetric part supplies an independent
quantum shell condition. The shell equation supplies information absent
from relational evolution by itself.

The spectral resolution of a fixed self-adjoint reduced
Hamiltonian yields an explicit distributional extension of
every reduced density operator supported on its negative spectral
subspace. For a
spectral dyad with energies $E_x$ and $E_y$, the reconstructed state is
supported at the clock momentum
$p_T=-(E_x+E_y)/2$ and evolves with the phase determined by
$E_x-E_y$. This pairing preserves off-diagonal coherences, satisfies the
left and right star constraints separately, and returns the reduced
Wigner state upon integration over $p_T$. The reconstruction is therefore
injective on the admissible sector.

The flat dust realization makes the spectral restriction concrete. With
$\alpha=k=0$, $q=1/2$, and Dirichlet data, the Hamiltonian is unitarily
equivalent to $(\hbar^2/2)d^2/dx^2$ on the half-line and has spectrum
$(-\infty,0]$. Hence its positive-density Schutz sector contains every
normalizable reduced state. The example also displays the boundary
condition and off-diagonal phases used by the reconstruction.

The scope remains tied to the deparametrized representative, ordering,
self-adjoint realization, half-line boundary data, and clock orientation.
The induced physical product is well defined on the distributional image;
its equality with an independently constructed group-averaged product is
an open comparison.

The reconstruction mechanism relies primarily on the linear form
$\mathcal C=p_T+H$ and on the spectral compatibility between the chosen
clock orientation and the reduced Hamiltonian. The same strategy can
therefore be tested in other deparametrizable cosmological models.
Other equations of state, spatial curvatures, self-adjoint realizations,
and internal clocks will determine how widely the admissible sector and
its reconstruction persist.

\section*{Data availability statement}

No new data were created or analysed in this study.

\bibliographystyle{unsrt}
\bibliography{referencias}

\end{document}